\documentclass[final,5p,times,twocolumn]{elsarticle}

\usepackage{graphicx}
\usepackage{caption}
\usepackage{subcaption}

\usepackage{amssymb}

 \usepackage{lineno}
\usepackage{natbib}
\usepackage[utf8x]{inputenc}

 \biboptions{numbers}

\journal{Biomedical Signal Processing and Control}

\begin{document}

\begin{frontmatter}

\title{Higher order spectral analysis of ECG signals}  

\author[label1]{Yamini Kotriwar}
\address[label1]{Indian Institute of Science Education and Research (IISER) Pune, 411008, India}
\address[label2]{Department of Physics, The Cochin College, Cochin 682002, India}
\address[label3]{Indian Institute of Science Education and Research Tirupati (IISER), 517507, India \fnref{label4}}
\cortext[cor1]{Correspondence and request of materials should be addressed to G.Ambika : IISER Pune, IISER Tirupati}

\ead{yamini.kotriwar@students.iiserpune.ac.in}

\author[label1]{Sneha Kachhara}

\ead{sneha.kachhara@students.iiserpune.ac.in}

\author[label2]{K. P. Harikrishnan}
\ead{kp\_hk2002@yahoo.co.in}

\author[label1,label3]{G. Ambika\corref{cor1}}

\ead{g.ambika@iiserpune.ac.in}
\ead{g.ambika@iisertirupati.ac.in}

\begin{abstract}
Higher Order Spectral (HOS) analysis is often applied effectively to analyze many bio-medical signals to detect nonlinear and non-Gaussian processes. One of the most basic HOS methods is the bispectral estimation, which extracts the degree of quadratic phase coupling between individual frequency components of a nonlinear signal. Most of the studies in this direction as applied to ECG signals are on the conventional, long duration (up to 24 hours) Heart Rate Variability (HRV) data. We report results of our studies on short duration ECG data of 60 seconds using power spectral and bispectral parameters. We analyze 60 healthy cases and 60 cases of patients diagnosed with four different heart diseases, Bundle Branch Block, Cardiomyopathy, Dysrhythmia and Myocardial Infarction. From the power spectra of these data sets we observe that the pulse frequency around 1 Hz has maximum power for all normal ECG data while in all disease cases, the power in the pulse frequency is suppressed and gets distributed among higher frequencies.  The bicoherence indices computed show that the pulse frequency has strong quadratic phase coupling with a large number of higher frequencies in healthy cases indicating nonlinearity in the underlying dynamical processes. The loss or decrease of the phase coupling with pulse frequency is a clear indicator of abnormal conditions. In specific cases, bicoherence studies coupled with spectral filter, suggest ECG for Myocardial Infarction has noisy components while in Dysrhythmia, power is mostly at high frequencies with strong quadratic coupling indicating much more irregularity and complexity than normal ECG signals. In addition to serving as indicators suggestive of abnormal conditions of the heart, the detailed analysis presented can lead to a wholistic understanding of normal heart dynamics and its variations during onset of diseases.    
\end{abstract}


\end{frontmatter}

\section{Introduction}
\label{sec1}
The electrocardiogram (ECG) records the electrical activity of the heart and is the externally available source of information on the pattern of spread of tiny electrical impulses produced by heart through the heart muscles \cite{waller1887demonstration}. Any variations in this pattern should reflect as subtle variations in the waveform of the ECG. Hence studies on the details of the waveforms of ECG can lead to very effective clinical diagnosis of abnormal functions of heart that are commonly called as heart diseases \cite{acharya2004classification}. The ability to interpret and evaluate different ECG signals is an important skill required for cardiac health care professionals. In this context, methods other than visual inspection, can definitely give an accurate and conclusive lead in diagnostics. In addition, study of ECG waveform can provide an understanding of the dynamics underlying normal heart rhythms and its variations during onset of abnormal conditions.

One of the most commonly analyzed measures is the Heart Rate Variability (HRV), which measures the time change in successive heartbeats \cite{abildstrom2003heart} and thus integrates responses from other physiological systems like nervous system, cardiovascular system, and respiratory system \cite{chuduc2013review}. Hence, HRV analysis plays a significant role in analyzing diseases of the heart. However, to gain a good understanding of the baseline HRV, ECG has to be recorded for a long duration, usually for hours. In contrast, the 12 lead ECG data used in normal clinical check ups is only 1-minute long. The analysis on the later, as reported here,  thus offers more accessibility and convenience over HRV and may even reflect irregularities missed by HRV analysis.

Each beat in the ECG comprises a P-QRS-T complex \cite{wang2007analysis,hurst1998naming} which is the time and amplitude representation of the signal. The ECG is analyzed in both time and frequency domain through different linear and nonlinear methods \cite{owis2002study,shekatkar2017detecting}. The time domain methods are usually filtering methods like auto regressive \cite{dantas2012spectral}  or distributed lag methods to assess cardiac autonomic activity \cite{ue2000assessment}. S{\"o}rnmo, Leif, et al.(1998) \cite{sornmo1998beat} had showed that beat-to-beat variability does not have sufficient information to discriminate between the presence or absence of coronary artery disease. Recently Sarusi et al. (2014) \cite{sarusi2014absolute} introduced a novel method of beat-to-beat variability and instability parameters and predicted ischemia induced ventricular fibrillation in rats' HRV. 

We start with the hypothesis that the subtle variations in the P-QRS-T waveform among beats will reflect as variations in the component frequencies and the dynamical couplings among them. Hence their power spectra combined with higher order spectral analysis could give first level information on the underlying dynamics as well as its variations due to pathological processes or absence of efficient feedback mechanisms. Since our methods use only the 60 seconds long 12 lead ECG data used in normal clinical check ups, the study will be attractive and useful for easy diagnosis. This could then be followed up with finer levels of analyses, as and when required.  

Over the last few decades, power spectrum has been widely used in the analysis of HRV and R-R peak intervals to detect arrhythmias \cite{bigger1992frequency,abildstrom2003heart,evrengul2005time} \cite{hunt2002t,ue2000assessment,minami1999real,stoica2005spectral}. Minami et al. (1999) \cite{minami1999real} had proposed real-time discrimination of ventricular tachyarrhythmia with Fourier transform neural network and were able to discriminate the ventricular rhythms from supra-ventricular ones. Evreng{\"u}l, Harun, et al. (2005) \cite{evrengul2005time} found through spectral analysis, that the patients with epilepsy show reduction in high frequency and increase in low frequency values, which may lead to ventricular tachyarrhythmia causing higher incidence of sudden deaths in epilepsy.  Ue, Hidetoshi, et al. (2006) \cite{ue2000assessment} did the assessment of cardiac autonomic nervous activities by means of ECG R-R interval power spectral analysis and successfully developed a computer system to study and discriminate between patients with ischemic heart disease and with varying degree of diabetic autonomic neuropathy. Staniczenko et al. (2009) \cite{staniczenko2009rapidly} successfully developed an algorithm based on power spectrum which is useful for the identification of frequency disorder in arrhythmias. Classification of arrhythmias based on power spectrum of the HRV data is reported by Khazaee et al. (2010) \cite{khazaee2010classification}. Recently, Takahashi, Naomi, et al. (2017) \cite{takahashi2017validity} used the maximum entropy spectral estimation method for spectral analysis in the frequency domain. 

In our study, we analyze short duration ECG signals using Fast Fourier Transform (FFT) which is a simple non-parametric algorithm with high processing speed and is one of the simplest and efficient methods to calculate power spectrum \cite{welch1967use}. The power spectra of all the 6 precordial leads for 60 healthy cases and 60 patients are obtained using the FFT algorithm and the frequencies with their corresponding power are extracted for further analysis.

The power spectrum only provides information about the distribution of the signal's power among its individual frequency components and does not provide any information on the phase relations between them. Hence most often higher order spectral analysis is used in the analysis of nonlinear signals like EEG \cite{ademoglu1992quadratic}, ECG \cite{alliche2003higher} as well as for pattern recognition \cite{chandran1997pattern}, analysis of harmonic random processes \cite{chandran1991mean}, geophysical processes \cite{elgar1989statistics}, turbines, astrophysical processes \cite{george2017detecting} etc. The first higher order spectrum is the bispectrum which is the Fourier transform of third-order correlation of the signal (power spectrum is Fourier Transform of the second order correlation). It indicates the cross-correlation between two frequency components in a two-dimensional frequency plot and gives information about the phase coupling between frequencies at $f_1$, $f_2$ and $f_1$+$f_2$. The normalized form of bispectrum is called bicoherence and its measures can show detectable trends in differentiating data sets.

In recent years, higher order spectral analysis has been put to good use in understanding variations in heart rhythms. Khadra et al. (2005) \cite{khadra2005quantitative} reported difference in bicoherence values from HRV of normal and tachycardia patients, indicating less phase coupling between the frequencies during an occurrence of arrhythmia. Chua et al (2008) \cite{chua2008cardiac} statistically classified different arrhythmia classes from HRV signals, using phase entropy and bispectral entropies that are lower for normal subjects. Later research in this direction \cite{martis2013cardiac} showed the bispectrum pattern for five different types of beats and automated the classification of these beats using neural networks. 

In a recent study reported by our group (Shekatkar et al. (2017) \cite{shekatkar2017detecting}) the complexity of the heart dynamics is characterized through the fractal nature of its dynamics. We proposed a novel method to differentiate the healthy and unhealthy cases, based on the resulting multifractal parameters. It is found that the fractal complexity of the healthy heart is reduced due to any abnormality in its functions. 

We report higher order spectral analysis done on short duration(60 secs) ECG signals using data sets of healthy as well cases of four specific diseases. The data sets used in the study are preprocessed using the methods of detrending followed by compression. The power spectra of all the processed datasets are presented and the frequencies with significant power are characterized. We further estimate their bispectral parameters and bicoherence values of pulse frequency with other significant frequencies to bring out the nonlinearity and quadratic phase coupling among them. We focus mainly on the pulse frequency and its phase coupling with other frequencies and estimate the number of frequencies  that have significant bicoherence with the pulse frequency. These numbers, when subjected to the analysis of variance (ANOVA) test, show significant `p' values ($<$ 0.001). Finally to check the presence of noisy processes, we apply bicoherence filter using the main peak frequency that brings out relevant disease specific features of the spectra. 

\section{Preprocessing of Data}
\label{sec}

\subsection{Data sets of ECG signals}
For the study presented here, we have used the data sets of PTB database from the `Physionet' Source \cite{goldberger2000physiobank}. This database contains records of ECG along with their header files with detailed medical information on the subjects including age, gender, primary and secondary diagnosis etc. Based on this information we have chosen ECG recordings of 60 healthy volunteers from the age group of 22 to 81 and 15 data sets each from patients in the age group of 17 to 82, diagnosed with four different diseases like Myocardial Infarction, Cardiomyopathy, Bundle branch block and Dysrhythmia. Some of the patients have secondary diagnosis like Hypertension, Diabetes Milletus, Atrial Fibrillation, Ventricular Fibrillation, Coronary Artery disease, Tinnitus etc. Each record consists of 15 simultaneously measured signals: the conventional 12 leads (i, ii, iii, avr, avl, avf, v1, v2, v3, v4, v5, v6) together with the 3 Frank lead ECGs (vx, vy, vz), each signal digitized at 1000 samples per second. The precordial leads correspond to the electrodes placed directly on the chest and data from each lead offers unique information which cannot be derived mathematically from the other leads \cite{kligfield2007recommendations}. We use data from the precordial leads from v1 to v6, also termed as `channels 1-6' for the total 120 cases considered in our analysis. 

\subsubsection{Detrending}
\begin{figure}[ht]
    \centering
    \includegraphics[width=\columnwidth]{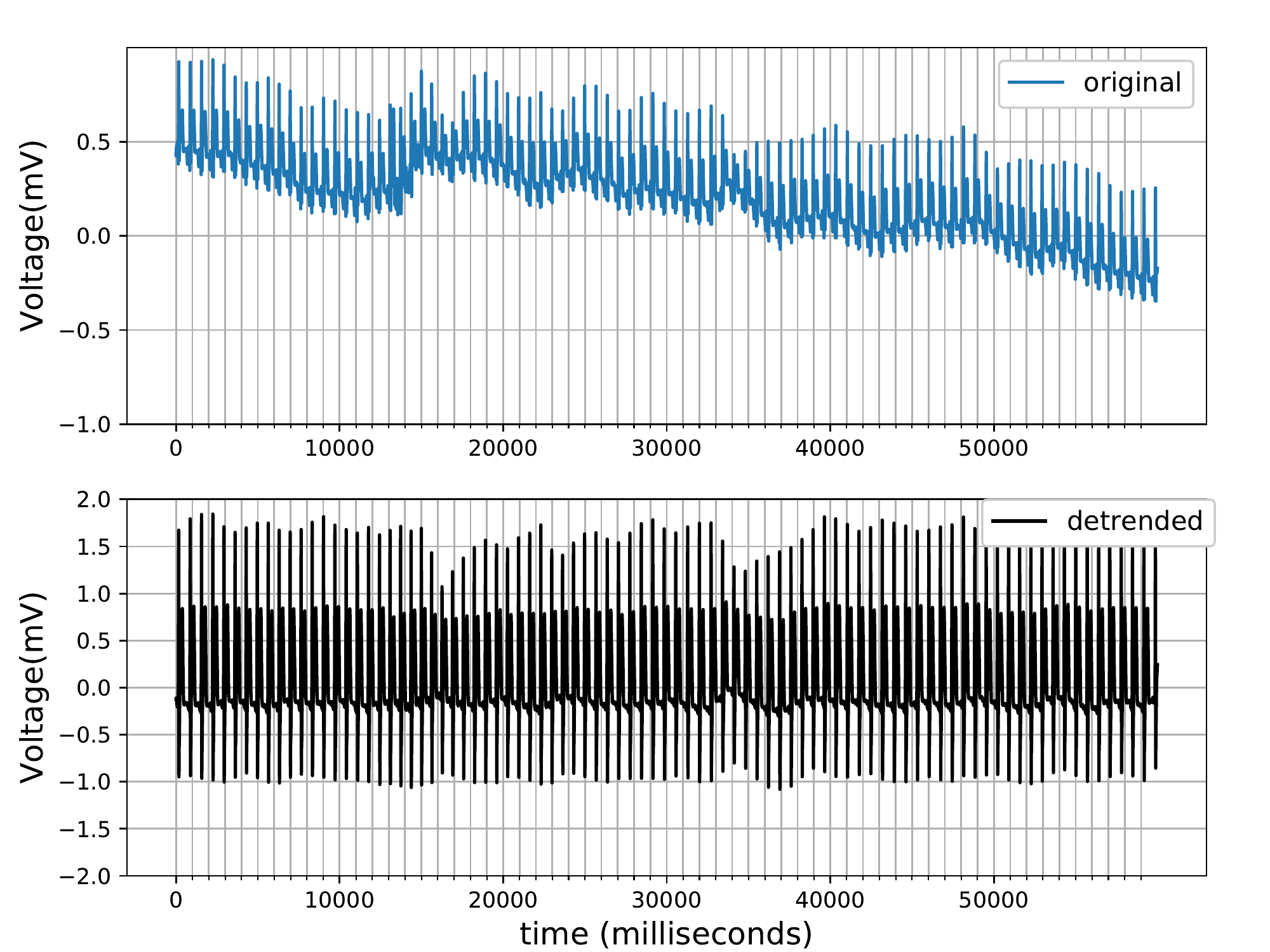}
    \caption{Time series of a typical ECG data set, before detrending (top) and after detrending (bottom). As explained in the text, the trend is removed by using the method of polynomial fitting.}
    \label{fig:detrend}
\end{figure}

Most of the ECG data sets indicate some global trends that can come from muscle movements of subject and/or the voltage fluctuations in the recording machine and hence needs to be detrended before further processing. This is achieved by using a polynomial fitting technique. For this to be effective, it is important to choose a suitable degree for the polynomial to fit the global trend, whose value is then subtracted from the original signal to get the detrended signal. The deviation $\delta_{(n)}$ of original signal $x_{(0)}$ from the detrended signal $x_{(n)}$, is defined as:
\begin{equation}
   \delta_{(n)} = \frac{1}{k}\sum_{i=1}^{k}(x_{(0)}(i)-x_{(n)}(i))^{2}
\end{equation}
where $k$  is the length of the signals. This deviation is calculated for different degrees of the polynomial, $n$. We find in all cases studied, $\delta_{(n)}$ saturates for values of $n$ higher than 20. Hence we take $n$ = 20 as the value of highest degree of the polynomial for all datasets to implement the detrending. The effect of detrending on a sample ECG signal is shown in the Fig. \ref{fig:detrend}. 

\subsubsection{Compression}
\begin{figure}[ht]
    \centering
    \includegraphics[width=\columnwidth]{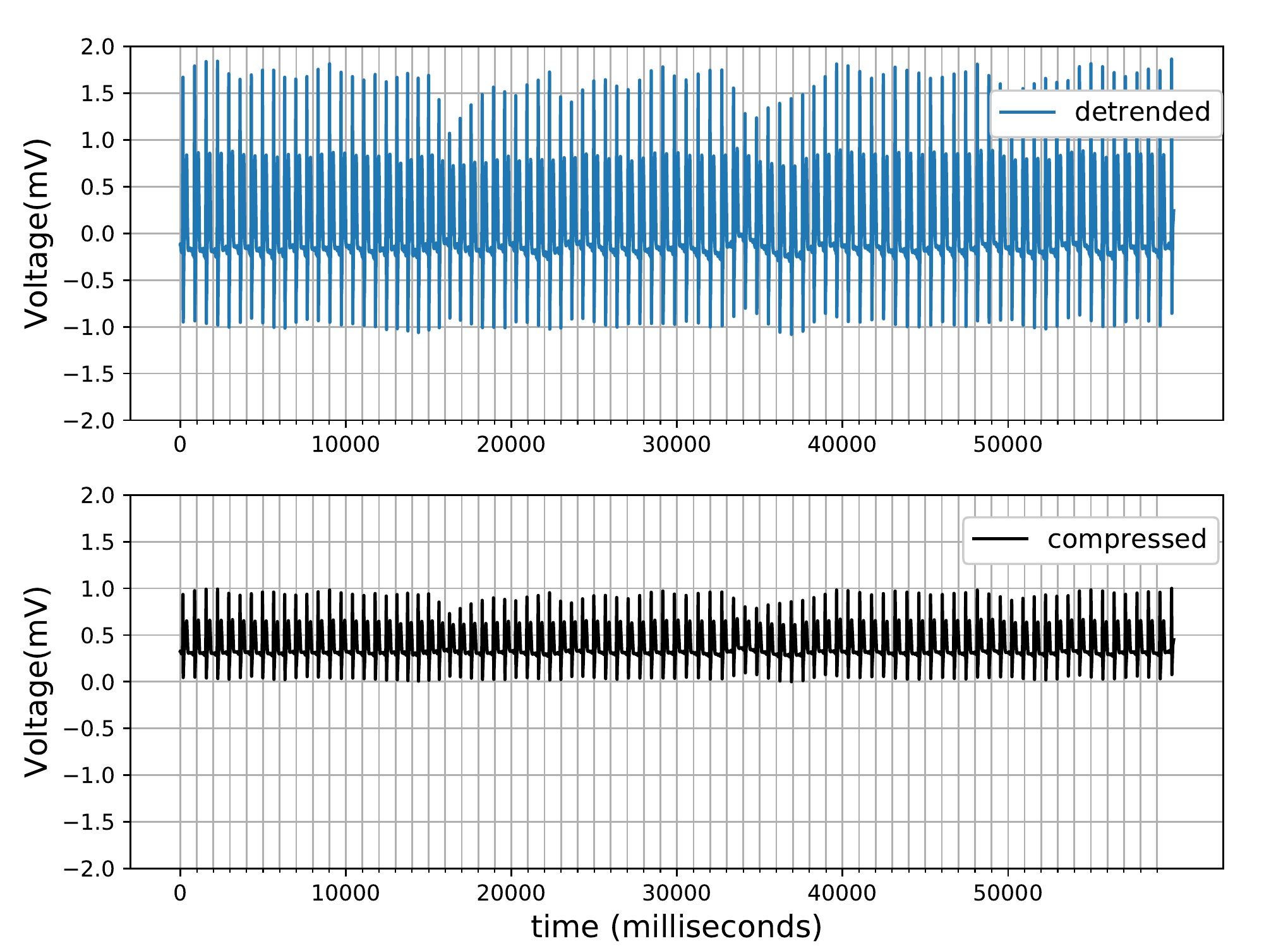}
    \caption{Time series of one sample detrended ECG data (top) and its compressed form (bottom). It is claer that the subtle trends and variations in the signal are preserved even after compression.}
    \label{fig:compressed}
 \end{figure}

It is necessary to bring all the signals into a uniform range for unbiased comparison before analysis. To achieve this uniformity for all the datasets, we used the method of ``compression", using:

\begin{equation}
s(t_{k})=\frac{c(t_{k})-x_{min}}{x_{max}-x_{min}}
\end{equation}

where $x_{max}$ and $x_{min}$ are the maximum and the minimum values of the ECG time series $x(t_{k})$ respectively and $c(t_{k})$ refers to the points in the compressed signal. The result of applying compression for one sample case is shown in the Fig. \ref{fig:compressed}. It is clear that this method sets all the values in (0,1) without altering the inherent information. 

\subsection{Power Spectral analysis}
\begin{figure}[ht]
    \centering
    \includegraphics[width=\columnwidth]{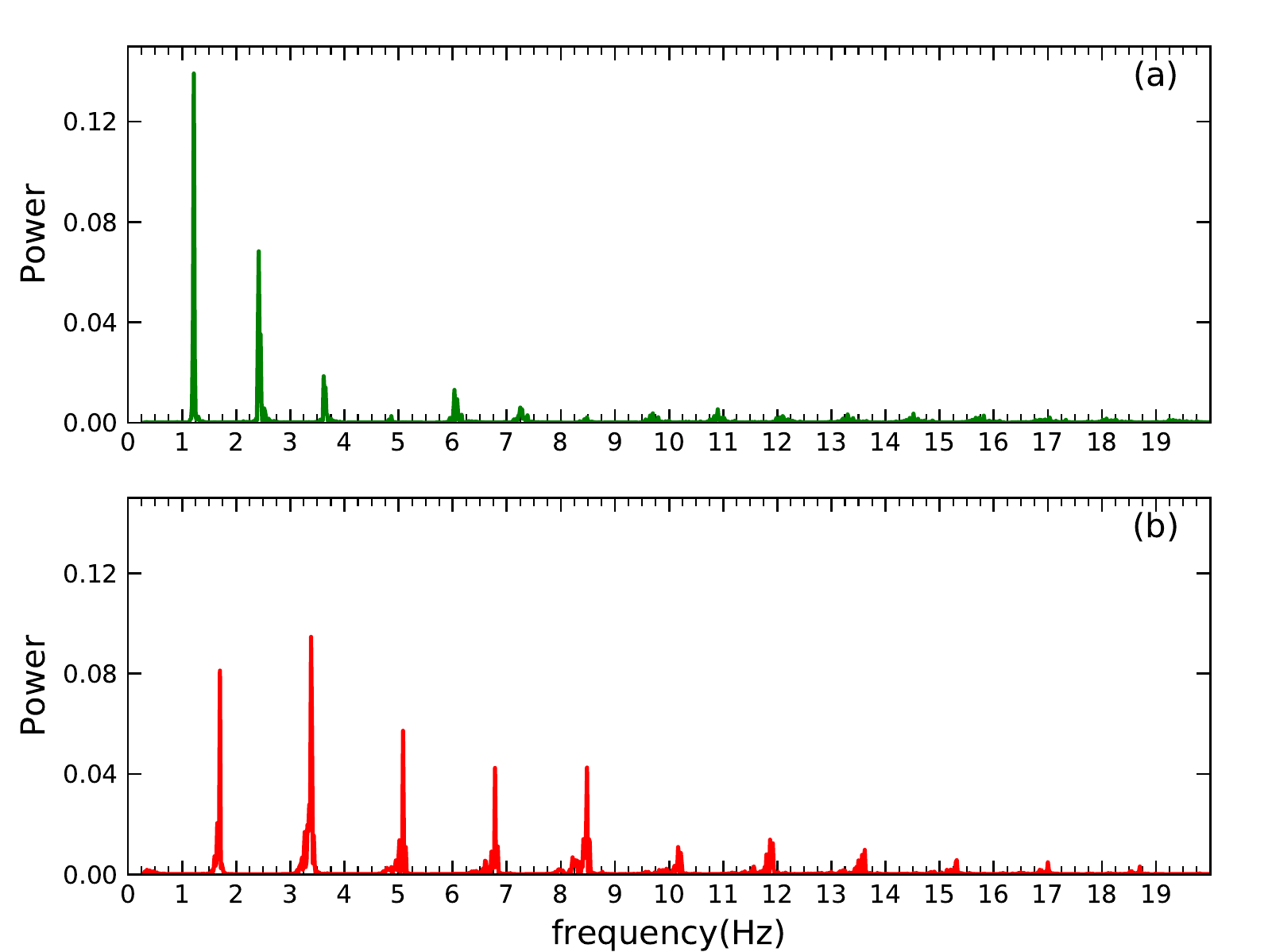}
    \caption{Power spectrum of (a) typical healthy ECG and (b) ECG of a patient with Myocardial Infarction. The distribution of power among higher frequencies is evident in the case of patient.}
    \label{fig:power_spectrum}
\end{figure}

For each data set we obtain the power spectrum using the Fast Fourier Transform (FFT) \cite{welch1967use}. The power spectra of one healthy and one disease case are shown in the Fig. \ref{fig:power_spectrum} as an illustration. In general, most of the power is distributed among the frequencies in the range of 0-20 Hz.

In an ECG signal, pulse frequency ranges from 1 Hz to 1.7 Hz, corresponding to the rhythms of normal heartbeat. We call this as the fundamental frequency of heart dynamics, $f_0$.  In healthy subjects, this frequency is the main component of the signal and should contain major fraction of the power in the power spectrum. To check this, we extract the frequency containing maximum power for each data (in all 6 channels) from the power spectrum and denaote it the `main peak frequency', $f_m$ of the power spectrum.

\begin{figure}[ht]
    \centering
    \includegraphics[width=\columnwidth]{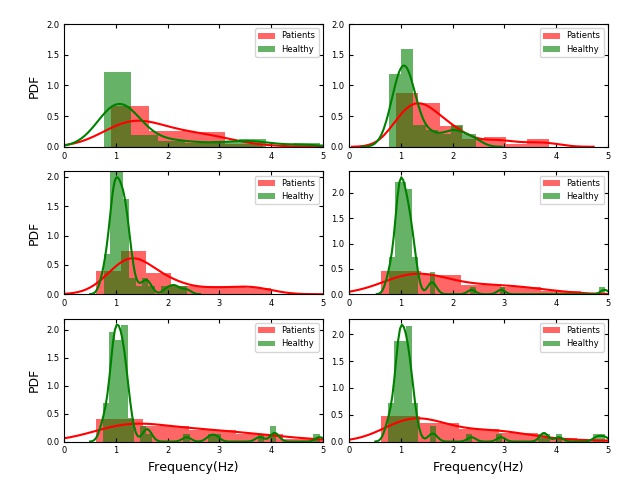}
    \caption{Distribution of first peak frequency($f_m$) of all the healthy and patients, for 6 channels. Note that the distribution is significantly wider in case of patients, most prominent in channels 5 and 6.}
    \label{fig:main_peak}
\end{figure}

The distributions of the main peak frequency $f_m$ for the healthy cases and patients are shown in the Fig. \ref{fig:main_peak} for all the 6 channels. (Here the frequencies less than 0.3 Hz are not included to avoid artifacts like baseline wander \cite{nayak2012filtering}) It can be seen that the distribution is centered around 1 Hz for healthy, whereas in case of patients it extends to higher frequencies, up to 4-5 Hz. The mean of the distribution for patients is shifted to the right, towards higher values of frequency. This analysis indicates that in healthy subjects, pulse frequency, $f_0$, is getting maximum power as expected. However, in patients, the power at pulse frequency is significantly reduced and a large fraction of power is redistributed among higher frequencies. 

As we know, ECG waveform can be thought of as a composite signal with many frequency components in addition to the pulse frequency, that arise from various physiological processes. Hence any abnormality in these processes could result in variations in the distribution of power among the component frequencies. Thus suppression of power in pulse frequency and redistribution among other frequencies can be taken as an indicator of any underlying abnormal conditions affecting the heart. 

\begin{figure}[ht]
    \centering
    \includegraphics[width=\columnwidth]{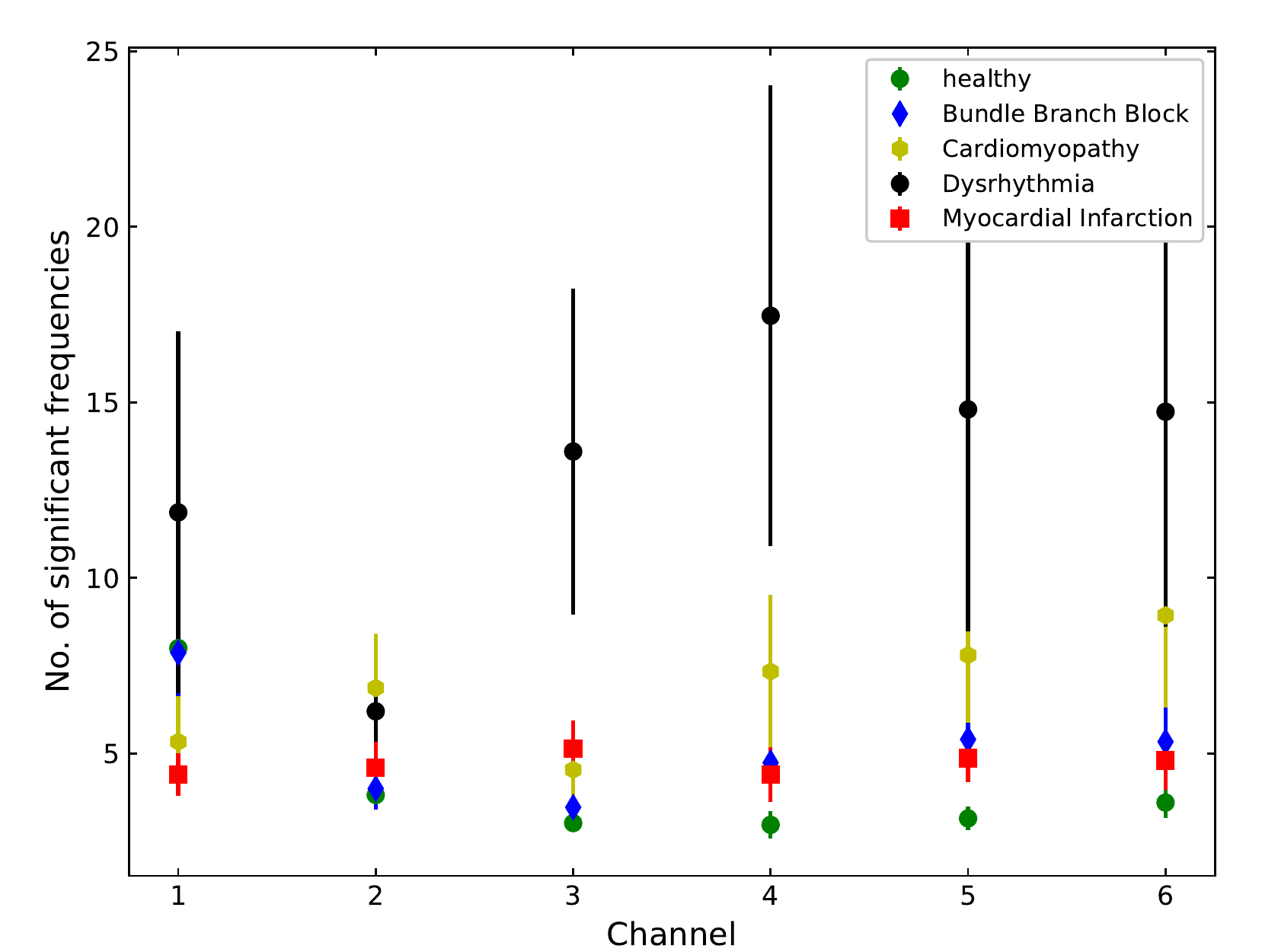}
    \caption{Average count of frequencies with more than 40\% of the maximum power for healthy and disease cases, for all the 6 channels. Compared to healthy, all diseases show higher number of significant frequencies with channel-to-channel variations, the highest being in the case of Dysrhythmia.}
    \label{fig:sig_freq}
\end{figure}

As described above, the power spectrum measures the power in different frequency components that make up the ECG signal. As evident from the main peak analysis, the distribution of power in cases of patients extends to higher frequencies with the corresponding power spectra showing a greater number of peaks (Fig. \ref{fig:main_peak}). We quantify this by counting the number of peaks with power above a threshold, set as  40\% of the maximum power. The count of such significant frequencies, with power more than 40\% of the maximum, averaged over each set of data is shown in Fig. \ref{fig:sig_freq}. All disease cases show larger values than normal cases, the highest being in the case of Dysrhythmia. Also channel-to-channel variations are seen to be much more evident for disease cases. The origin of these additional frequencies (as compared to healthy) can be traced to the irregularities in the heart rhythms, which indicates reduction in the underlying self-regulatory mechanisms that keep a normal heart beating in a certain regular pattern.

As is clear from Fig. \ref{fig:main_peak} and Fig. \ref{fig:sig_freq}, in case of healthy, the heart beats are variable and the signal is aperiodic but the extent of this variability is less \cite{dekker2000low}. The dynamics of a healthy heart leads to a more or less limited range of pulse rate (60-100 beats per minute for most adults) and has feedback from itself on top of other physiological processes like respiration etc. The underlying polarization-depolarization cycle of the cardiac muscles is more or less regular and thus results in a limited number of significant frequencies. In contrast, cardiac abnormalities hinder the natural cycle and result in higher variability in the ECG signals. These changes are most pronounced during Dysrhythmia with a large number of frequency components sharing significant power. Thus presence of significant power in higher frequencies that may or may not be the overtones of the pulse frequency, though not conclusive at this stage, is suggestive of abnormalities in the heart beat mechanisms.

\subsection{Higher Order Spectral Analysis}
From the power spectral analysis, we find that there are many higher frequencies other than the pulse frequency and its harmonics, which are getting significant power. Since power spectrum is a linear analysis, it cannot give any further understanding on the role of these frequencies or dynamical relations among them. For this, we have to do higher order spectral analysis, using bispectra and bicoherence calculations. 
Bispectral estimation extracts the degree of quadratic phase coupling between individual frequency components of the signal. To calculate the bispectrum, the data is divided into $N$ number of segments and their Fourier transforms calculated \cite{rao2012introduction}. Then, bispectrum is given by

\begin{equation}
b(f_1,f_2)= \sum_{j=1}^{N}X_j(f_1) X_j(f_2)X_j{^*}(f_1+f_2)
    \end{equation}
where N is the number of segments, each with length 8192, in our calculations. X$(f_1)$ and X$(f_2)$ are the discrete-time Fourier transform computed as discrete Fourier transform (DFT) using the FFT algorithm. The bispectrum is then normalized so that is has a magnitude in the range (0,1) \cite{nikias1987bispectrum, nikias1993higher}, to get the corresponding bicoherence, as:

 \begin{equation}
   B(f_1,f_2)= \frac{|\sum_{j=1}^{N}X_j(f_1) X_j(f_2)X_j{^*}(f_1+f_2)|}{\sum_{j=1}^{N}|X_j(f_1) X_j(f_2)X_j{^*}(f_1+f_2)|}  
     \end{equation}

\begin{figure}[ht]
    \centering
    \includegraphics[width=\columnwidth]{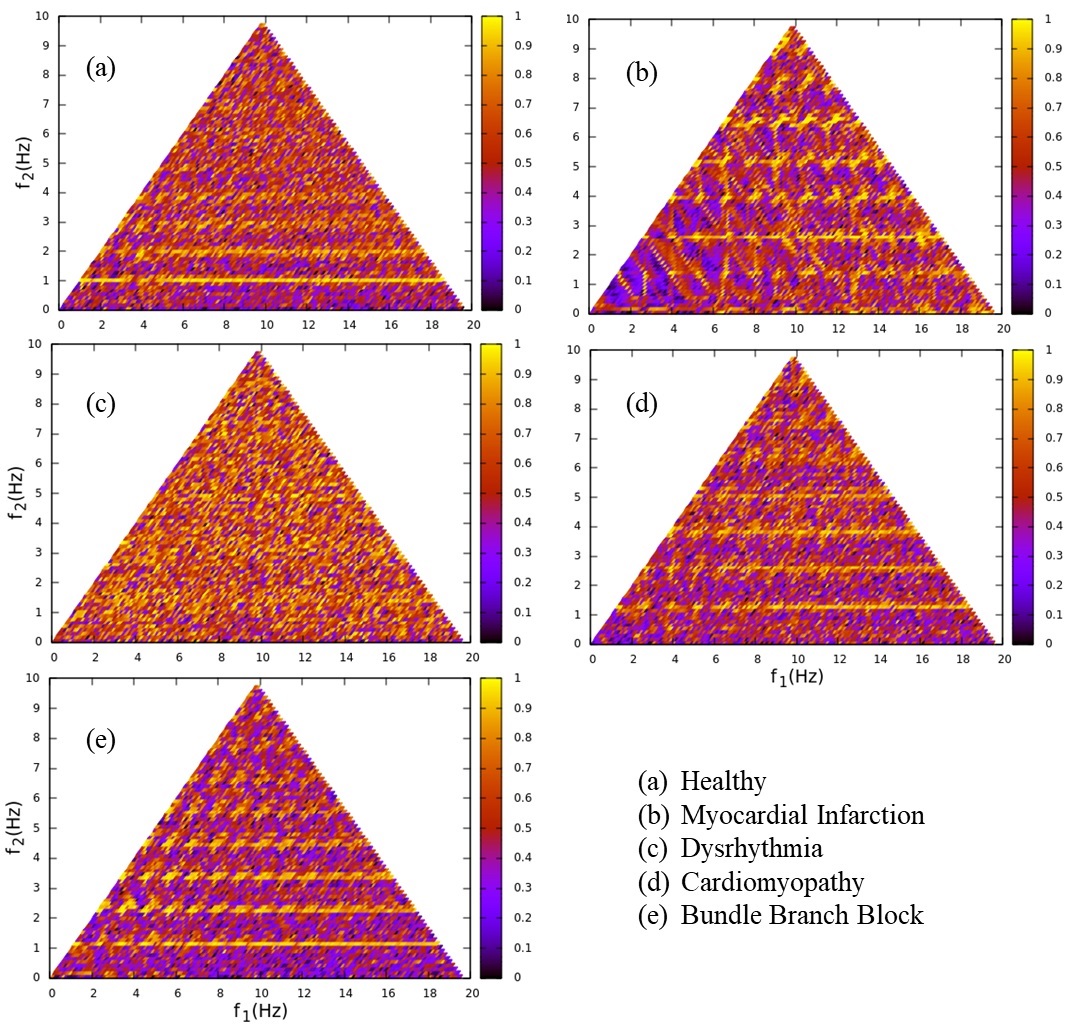}
    \caption{Bicoherence plots for (a) Healthy, (b) Myocardial Infarction, (c) Dysrhythmia, (d) Cardiomyopathy and (e) Bundle Branch Block. The pulse frequency(${\sim}1 Hz$) has significant bicoherence with its harmonics as well as other frequencies in the case of healthy and Dysrhythmia. However, the power in these frequencies is higher in Dysrhythmia as compared to healthy (as observed in their power spectra). In the case of Myocardial Infarction, the bicoherence is significant only for harmonics of the pulse frequency. Almost similar behavior is seen for Cardiomyopathy and Bundle Branch Block.}
    \label{fig:bic_plots}
\end{figure}

It is an auto-quantity, i.e. it can be computed from a single signal. The bicoherence measures the proportion of the signal energy at any frequency pair that is quadratically phase coupled \cite{totsky2015bispectral}.
Due to inherent symmetries, the bicoherence function of a real valued signal can be defined in the triangular region given by $f_2>0$ ; $f_1\geq f_2$  ; $f_1 + f_2  \leq f_{max}$ where $f_{max}$ is the Nyquist frequency (half of the sampling rate) of the data \cite{totsky2015bispectral}. Only those values of bicoherence are considered which are above 80\% significance level, given as $\sqrt{\frac{9.2}{2*N}}$. In the power spectra of ECG data sets used, we observe that most of the power is distributed over the frequencies in the range of 0-20 Hz. Hence the bicoherence is also calculated for the same range. The
bicoherence plots thus obtained for one sample case from each disease and one healthy case are shown in Fig. \ref{fig:bic_plots}.

\begin{figure}[!ht]
    \centering
    \includegraphics[width=\columnwidth]{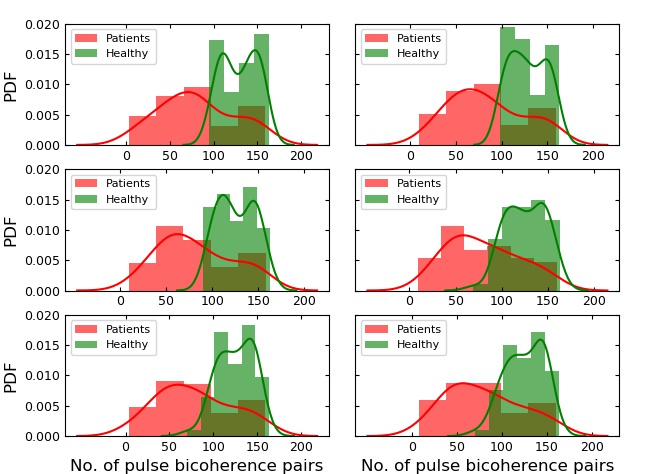}
    \caption{Distribution of the number of frequencies showing significant bicoherence with the pulse frequency ($f_{0}$) for healthy and patients. The number of pulse bicoherence pairs $n(f_{0},f_{i})$ in healthy is always more than patients in all the 6 channels.}
    \label{fig:pulse_bic_pairs}
\end{figure}

Considering the pulse frequency $f_{0}$ as the most fundamental one in heart rhythms, we calculate its bicoherence with other frequencies and the number of frequency pairs $f_{0},f_{i}$ with significant bicoherence (i.e. more than 80\%) is obtained for every channel for each data set. The distribution of this number, $n(f_{0},f_{i})$, is shown in Fig. \ref{fig:pulse_bic_pairs}. We observe that $n(f_{0},f_{i})$ is higher than 90 for all healthy subjects, while the distribution is wider for patients, ranging from 10 to 170. 

The distributions of $n(f_{0},f_{i})$ for healthy and disease sets can be further distinguished using ANOVA test. The Analysis of variance (ANOVA) is a statistical method that can be used to check if the distributions of multiple classes are significantly different from each other. In general, it provides a quantifying measure of the resemblance of two given distributions with each other. The ANOVA algorithm calculates a p-value for each class, and a lower value of p, indicates lower the probability to get a distribution similar to what is given, which means a high probability for the distributions to be  statistically different from each other \cite{neter1996applied}. We apply the ANOVA test on the distribution of pulse frequency bicoherence pairs $n(f_{0},f_{i})$ for the two classes, healthy and patients, we get p-values significantly lower than 0.001, for all the six channels, confirming that their distributions are very different.

We also calculate the number of frequency pairs $f_{m},f_{i}$ with significant bicoherence, $n(f_{m},f_{i})$ and the number of all frequency pairs other than these two, but with significant bicoherence as $n(f_{i},f_{j})$, for all cases. These give interesting disease specific indications from their relative values. As expected for all healthy cases, $n(f_{0},f_{i})$ is almost the same as $n(f_{m},f_{i})$ with large value for $n(f_{i},f_{j})$ also. This once again confirms the dominance of pulse frequency $f_{0}$ in healthy conditions, as well as the nonlinear and chaotic nature of its underlying dynamics.

In the case of Myocardial Infarction, $n(f_{m},f_{i})$ is greater than $n(f_{0},f_{i})$  with $n(f_{0},f_{i})$ being less than 100. For all other diseases, these two numbers are almost equal on the average but show larger variability from channel to channel. Also in the case of Dysrhythmia, the number of the remaining frequencies, $n(f_{i},f_{j})$, is high indicating many more frequencies that have dynamical origin. This would lead to the conclusion that dynamics of the heart under Dysrhythmia conditions is more chaotic than the healthy heart.

\subsection{Main peak bicoherence filter}

The large number of peaks seen in the power spectrum need not all be of dynamic origin and especially that do not have significant bicoherence can be of noisy origin. To confirm this and check the presence of noisy or stochastic nature in underlying processes, we apply the method of filtering using main peak bicoherence \cite{george2017detecting}. While our calculations so far are with the threshold for significant power fixed as more than 40\% of the maximum in the power spectra, we now vary this threshold and plot the total number of frequencies having power above a chosen threshold, $n_{i}$ as well as the number of frequencies that have significant bicoherence with the main peak $n_{b}$ in each data set.

\begin{figure}[ht]
    \centering
    \includegraphics[width=\columnwidth]{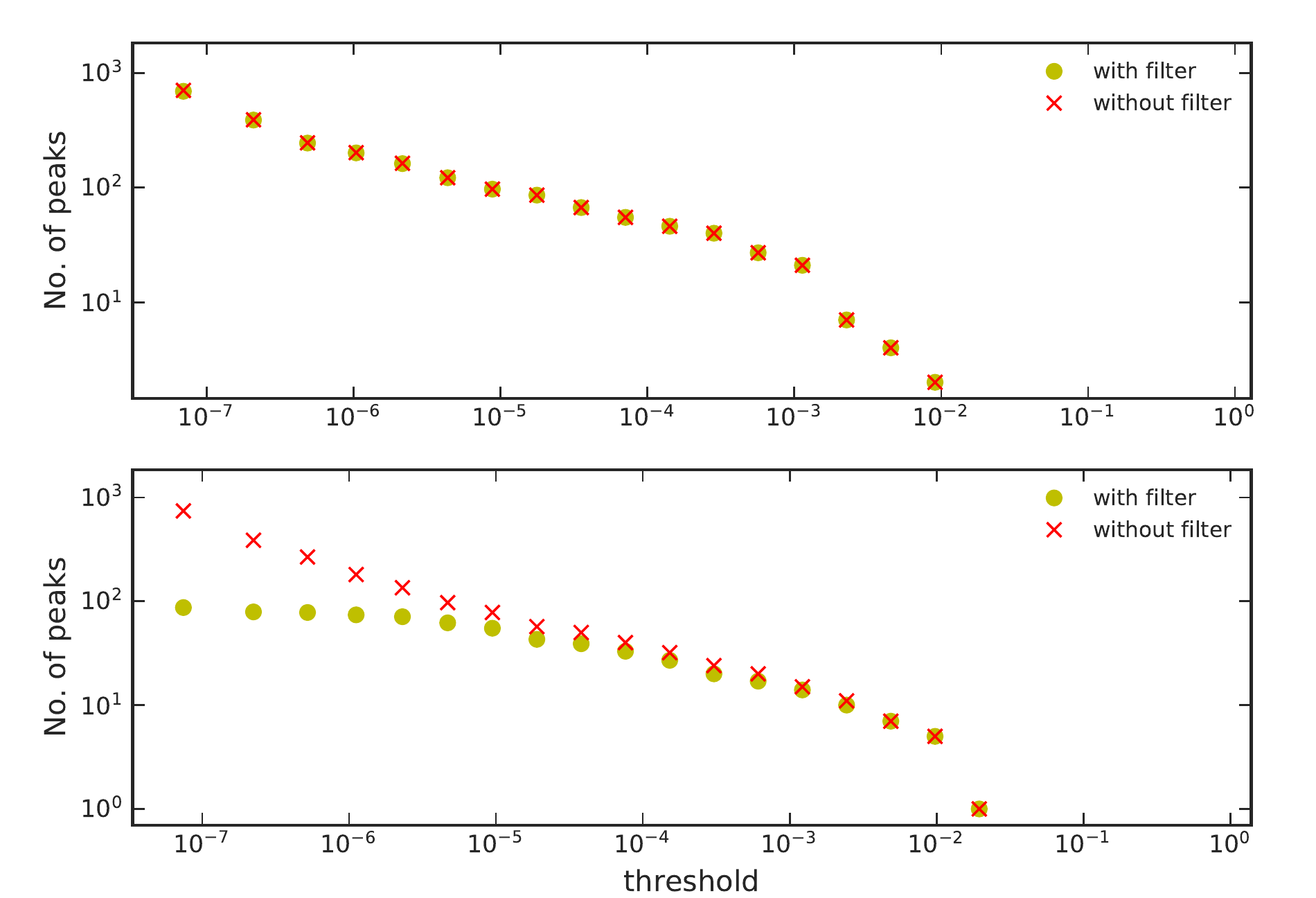}
    \caption{Application of main peak bicoherence filter for (a) a typical healthy data and (b) data for Myocardial Infarction. The counts remain the same for (a) with and without the bicoherence filter validating the result that the peaks in the power spectrum of healthy are of dynamical origin whereas in case of (b) the counts of peaks after using the bicoherence filter are reduced indicating presence of stochastic processes.}
    \label{fig:bic_filter}
\end{figure}

The frequencies included in $n_{i}$, but are not present in $n_{b}$ may not be due to a dynamical origin, and could arise from noise or other artifacts. 
Therefore as we vary the threshold, if these two numbers change for the same data, then it is an indication of noisy or stochastic origins. They may also be due to higher order phase coupling, although less likely in the present context.  
We analyze our data with the above filter and the results for two sample cases (one healthy and one patient with Myocardial Infarction) are shown in Fig. \ref{fig:bic_filter}. In the case of healthy, the filter does not change the numbers as the threshold is lowered, a strong evidence for the significance of all the frequencies and their dynamical origin. However, in the case of Myocardial Infarction, we see the number of peaks with filter to be significantly diverging from that of without filter, indicating that there are many frequencies that are not actually phase coupled to the main peak frequency and thus may be due to noisy origin. 

\section{Summary and Discussion}

The basic ECG signal contains information about the underlying dynamics of the heart. Human heart is understood to be a chaotic system \cite{garfinkel1997quasiperiodicity}, and the complex variations in waveform of ECG should reflect this nature. However, there is a regular beat pattern, which indicates the physiological self-regulatory feedback mechanisms. In the case of healthy heart, the signal has a specific shape in the time domain for its P-QRS-T waveform with heart beat rates of 60-100 beats per minute, corresponding to the pulse frequency of around 1 Hz. Ideally, this frequency $f_{0}$ should be the main peak with maximum power in the power spectrum. We consider this to be the most fundamental frequency and study its dynamical relationship with its harmonics and other frequencies. The results of spectral analysis on 60 data sets from healthy and 60 from four different disease cases, show that in almost all cases of healthy, the main peak in the power spectrum is centered around a narrow range of 1-2 Hz, but for patients it goes up to 4 Hz. This might indicate reduced efficiency in the feedback mechanism of the heart and can serve as a precursor to higher order methods to identify the cause of the erratic functions.
 
In addition to the main peak, there are frequencies that get a significant share of the power. We estimate the number of peaks that have power more than 40\% of the maximum and find this number to be low (less than 5) for healthy subjects. However for data from patients, it is always higher with a remarkable difference in the case of Dysrhythmia where it is in the range 10-20. We also observe channel-to-channel variations in this number for the later set.
For the healthy cases, maximum power is in the pulse frequency only with the power in higher frequencies being much less. During onset of heart problems, the power in $f_{0}$ is suppressed and redistributed to higher frequencies.
Thus for the typical case of Dysrhythmia, the power is much higher in the overtones. Our analysis shows that the general signature of the abnormality or disease is the presence of higher frequency components in the power spectrum that have significantly more power than $f_{0}$. These frequencies may or may not be the harmonics of the pulse frequency, depending on the dynamics of the diseased heart.
 
Hence to characterize the nature of the underlying conditions further, we have to check whether all the higher frequencies are dynamical in origin. For this, we do a comprehensive bispectral analysis and estimation of bicoherence indices from the spectra of ECG signals. 

While the pulse frequency indicates strong phase coupling with other frequencies in the healthy cases, the number of frequency pairs with significant bicoherence with the pulse frequency, $n(f_{0},f_{i})$, indicates disease specific variations.  We also calculate the number of frequency pairs $f_{m},f_{i}$ with significant bicoherence, as $n(f_{m},f_{i})$ and the number of all frequency pairs other than these two, but with significant bicoherence as $n(f_{i},f_{j})$, for all the data sets studied. 

For all healthy cases, $n(f_{0},f_{i})$ is almost the same as $n(f_{m},f_{i})$. 
In the case of Myocardial Infarction, $n(f_{m},f_{i})$ is greater than $n(f_{0},f_{i})$. For all other diseases, these two numbers are almost equal on the average but show larger variability from channel-to-channel. The values of $n(f_{i},f_{j})$ are almost the same as healthy for most of the diseases but again with channel-to-channel variations, except for the case of Dysrhythmia. For Dysrhythmia $n(f_{i},f_{j})$, is higher indicating many more frequencies that are of dynamical origin. Having large values for $n(f_{0},f_{i})$ or $n(f_{m},f_{i})$ indicates the nonlinear nature of the underlying dynamics while large number of extra pairs with significant bicoherence $n(f_{i},f_{j})$ indicates chaotic nature in the underlying dynamics. Hence our results are suggestive of the fact that while heart dynamics is in general nonlinear and chaotic, its dynamics under Dysrhythmia conditions is more chaotic than the healthy heart.

In this context we note that Garfinkel et al.(1997) had provided the first experimental validation of the multi frequency quasi periodic transitions leading to a spatio-temporal chaos in cardiac fibrillation. Their analysis suggest that there are frequencies that have a dynamical origin resulting in chaos which agrees well with our findings in the cases of Dysrhythmia, that have fibrillation as secondary diagnosis \cite{garfinkel1997quasiperiodicity}. 
Myocardial Infarction is typically hard to detect at an early stage \cite{mendis2010world}, however our bicoherence analysis shows a clear difference in the range of the number of pulse frequency bicoherence pairs. 

To isolate peaks in the power spectrum that do not have dynamical origin but may have noisy or random origins, we apply the method of bicoherence filter using the main peak frequency. We observe that in healthy cases, the number of peaks with and without bicoherence filter is the same even at lower values of threshold for power, which indicates that all these frequencies are of dynamical origin. Since in this case pulse frequency gets maximum power, we can say all frequencies have a strong phase coupling with the pulse frequency. However, in cases of Myocardial Infarction, application of filter indicates a reduction in the number of peaks for low values of threshold. This reduction indicates that the filtered out frequencies may have stochastic or noisy origins, which is not observed in other diseases or healthy heart.\\
The main results of significance from our study are

For healthy cases, the maximum power is at the pulse frequency and it has significant bicoherence with its harmonics and other higher frequencies indicating nonlnearity and chaos in the underlying dynamics.

The power at pulse frequency is suppressed and redistributed to higher frequencies during any malfunctions of this normal dynamics. 

There is an increase in the number of frequencies having significant power in the spectra for diseases.
 
In general variations from normal behaviour are more pronounced in channels from 3-6.

The number of frequency pairs having significant bicoherence with pulse frequency is less than that with main peak in the case of Myocardial Infarction which also reveals peaks with noisy origin in the power spectra.

During Dysrhythmia, the total number of significant peaks in power spectrum is much larger. They have significant bicoherence with pulse frequency, main peak frequency and a set of other frequency indicating stronger chaotic nature than dynamics of normal heart. 

The present study was done only on a limited number of cases and some of the patients had secondary diagnosis as well which obscures or masks the specific trends, making it difficult to identify a disease-specific feature. Because of this, our analysis reported here, could not identify a common classifier for Cardiomyopathy and Bundle branch block. But we get identifiable features for Myocardial Infarction and Dysrhythmia in their power spectrum and bicoherence.

In conclusion, higher order spectral analysis coupled with power spectra is found to be a good potential indicator for the classification of cardiac abnormalities. The fact that it is done with the short duration ECG data makes it useful in clinical practice. So also as it reduces computation time and can be automated. In general our results on the distribution of pulse frequency and the number of frequency pairs with significant bicoherence together can help in differentiating healthy cases from the patients. At this stage, our findings can be suggestive of any abnormalities that need medical attention. A detailed investigation using more data sets along similar lines on the significant frequencies in the higher order spectrum and their role in the heart dynamics could be the focus of further research.

\section*{Acknowledgement}
The authors acknowledge the financial support from Dept. of Sci. and Tech., Govt. of India, through a Research
Grant No. EMR/2014/000876. We also acknowledge the Physiobank data base (www.physionet.org/physiobank/database/) for the ECG data used in the study reported here.

\bibliographystyle{elsarticle-num-names}

\end{document}